\documentclass[12pt]{article}  
\usepackage{a4,epsfig,graphics,subfigure}
\begin{document}




\begin{titlepage}
\begin{center}
\vspace{-0.5cm}
{\Large EUROPEAN ORGANIZATION FOR NUCLEAR RESEARCH}
\end{center}
\begin{flushright}
  LHWG note 2001-07 \\
  ALEPH  2001-058 PHYSICS 2001-038\\
  DELPHI 2001-118 CONF 541 \\
  L3 Note 2703 \\
  OPAL Technical Note TN700\\
  5 July, 2001 \\
\end{flushright}
\bigskip
\vspace*{1.5cm}
\begin{center}{\Large \bf Flavour Independent Search for Hadronically 
Decaying 
Neutral Higgs Bosons at LEP} \\
 \bigskip
      {\large The LEP Working Group for Higgs Boson Searches}
\end{center}
\vspace*{1.5cm}
\bigskip
\begin{center}{\Large  Abstract}
\end{center}
Hadronic decays of Higgs bosons, not necessarily into $b$-quarks, have
been searched for using data collected at LEP-2. Such searches are
complementary to the usual Standard Model Higgs searches and lead to
more model-independence. Preliminary results obtained by the four LEP
collaborations are presented for the $hZ$ production mechanism (using
the $q\bar{q}q\bar{q}$, $q\bar{q}\nu\bar{\nu}$ and $q\bar{q}l^+l^-$
channels) and combined for the first time.
\vspace*{3.5cm}
\begin{center}
ALL RESULTS QUOTED IN THIS NOTE ARE PRELIMINARY\\
(contributed paper for EPS'01 in Budapest and LP'01 in Rome)
\end{center}
\end{titlepage}
\pagebreak

\setcounter{page}{1}    



\section{Introduction}

There are extensions of the Standard Model in which Higgs bosons
have suppressed couplings into $b$-quarks. 
This can occur for specific parameters of the Two Higgs Doublet 
Model \cite{2HDM-flavind}, or of the Minimal SuperSymmetric Model 
\cite{MSSM-flavind}, as well as for some composite models \cite{calmet}. 
Standard Model Higgs searches~\cite{LEP-smhiggs} would have a reduced
sensitivity in such cases, because of their strong reliance on the 
identification of the $b$-quarks from the Higgs boson decay to maximize the
separating power. It is important to also cover such scenarios 
experimentally with dedicated searches in which the information
from the flavour of the quarks in the Higgs boson decay is not exploited,
so that the model-dependance of our final Higgs results can be reduced.

All four LEP collaborations have pursued such {\it flavour independent} 
searches in recent years~\cite{flbl-A,flbl-D,flbl-L,flbl-O},
analysing the four-jet ($q\bar{q}q\bar{q}$), missing energy 
($q\bar{q}\nu\bar{\nu}$) and leptonic ($q\bar{q}l^+l^-$) topologies.
None has found evidence for any signal.
In this note, the first combination of the results obtained is presented, for
the $e^+e^- \rightarrow Z^* \rightarrow hZ$ production mechanism, in
terms of upper limits on the corresponding cross-section
as a function of the Higgs boson mass, and of a lower mass limit
in the assumption of production cross-sections equal to those in the
Standard Model and a Higgs boson decaying into hadrons with a 
100 \% branching fraction. Higgs mass assumptions from 60 to
115 GeV were tested.

Some systematic effects have been studied by individual experiments,
but were not considered nor included in the evaluation presented here. 
All results are preliminary.

\section{Experimental Analyses}

The analyses used by the four collaborations to 
search for $hZ$ production in the flavour independent hypothesis
were to a large extent adaptations of existing LEP-2 searches
or measurements. 
In all channels except four-jets, events were selected
exactly, or almost exactly, as in 
the corresponding Standard Model Higgs search, removing 
the $b$-tagging from the final selection. In the four-jet channel, on
the other hand, all collaborations have used dedicated test-mass
dependent selections, to exploit maximally the kinematic features 
available and the mass reconstruction. This was necessary 
to enable reducing as much as possible the dominant backgrounds from 
$WW$ and $ZZ$ production and from QCD processes $qqgg$, 
in the absence of $b$-tagging.

Because the searches performed by the four collaborations have
been developed relatively, or in some cases 
very, recently, they are less optimised,
slightly less comprehensive, and
have used more simplifying assumptions to extract the results
than the corresponding searches for the Standard Model Higgs boson.

In the ALEPH search~\cite{flbl-A}, the data collected 
in 1998, 1999 and 2000 were used, clustering the data from the last year
in seven energy bins. All the topologies corresponding
to possible decay products of the $Z$ boson were investigated. In the four-jet
channel, a dedicated search based on a neural network method was
developed. The compatibility of the data with the signal hypothesis was 
tested in the Higgs boson mass range from 60 up to 115 GeV/c$^2$, 
with 5 GeV/c$^2$ steps, interpolating the final discriminant variable
to test intermediate values.
 
In the DELPHI search~\cite{flbl-D}, only the data collected
in 1999 and 2000 were used, clustering the data from the last year
in two energy bins. All the topologies corresponding
to possible decay products of the $Z$ boson were investigated,
except the $q\bar{q}\tau^+\tau^-$ final state.
The compatibility of the data with the signal hypothesis was 
tested in the Higgs boson mass range from 50 up to 110 GeV/c$^2$, 
with 5 GeV/c$^2$ steps. A small degradation of the performance 
was introduced in the final evaluation, between each test mass, 
to account for mass resolution effects.

In the L3 search~\cite{flbl-L}, only the data collected 
in 1999 and 2000 were used, clustering the data from the last year
in five energy bins. All the topologies corresponding
to possible decay products of the $Z$ boson were investigated. 
The compatibility of the data with the signal hypothesis was 
tested in the Higgs boson mass range from 60 up to 115 GeV/c$^2$, 
with 1 GeV/c$^2$ steps.

In the OPAL search~\cite{flbl-O}, the data collected 
in 1998, 1999 and 2000 were used, clustering the data from the last year
in a single energy bin. Dedicated analyses were conducted for
all the topologies corresponding to possible decay products of the $Z$ boson.
For the four-jet channel the test-masses were chosen in the Higgs boson
mass range from 60 up to 115 GeV/c$^2$, with 1 GeV/c$^2$ steps.

Finally, in spite of not taking advantage of the $b$-tagging of the jets from
the Higgs boson decay, every experiment has found small 
but still significant differences in performance
between the different possible decay products of the Higgs boson, 
arising from slight differences in mass resolution and jet structure. 
In particular, Higgs boson decays into gluon pairs have larger
multiplicities, but at the same time coarser dijet mass resolution, 
than decays into light quarks.
In order to enable quoting genuine flavour independent results, the samples used for 
the final evaluation were conservatively chosen as those 
which gave the weakest expected performance in each channel, and 
for each value of the Higgs boson mass.

\section{Results}

The common evaluation of the results used the standard statistical 
procedures based on the likelihood ratio technique, as applied in the other
combinations performed by the LEP working group for Higgs boson 
searches \cite{lephwg-stat,LEP-smhiggs}. 

Results were first obtained for each of the four collaborations, 
with two independent implementations of the combination
software \cite{O-stat,L-stat}, and compared with the results obtained 
within each collaboration.
Although some slight differences were found, the general shapes and features
of the observed and expected confidence levels for the signal, 
$CL_s$ and background-only, $CL_b$ hypotheses were very consistent. 

The calculations of the observed and expected 95\% CL lower limits on 
the mass of the Higgs boson, assuming production 
cross-sections equal to those in the Standard Model and
BR(h$\rightarrow$ hadrons) = 1.0, all agreed 
within a few hundred MeV, or better. The values\footnote{Because the
evaluations performed did not include systematic uncertainties, 
the values obtained differ in some cases from those 
quoted by the individual colloborations which included them in 
their evaluations. This is most notably the case for the OPAL 
results~\cite{flbl-O}.} from one of the implementations \cite{O-stat} 
are shown in Table \ref{leplim}, together with the combined results 
using the data from all four collaborations. 
The combined observed and median expected limits were 112.9 and 113.0 
GeV/c$^2$, respectively.

\begin{figure}[tbh]
  \begin{center}
   \mbox{\epsfxsize=13.0cm\epsfysize=10.0cm\epsffile{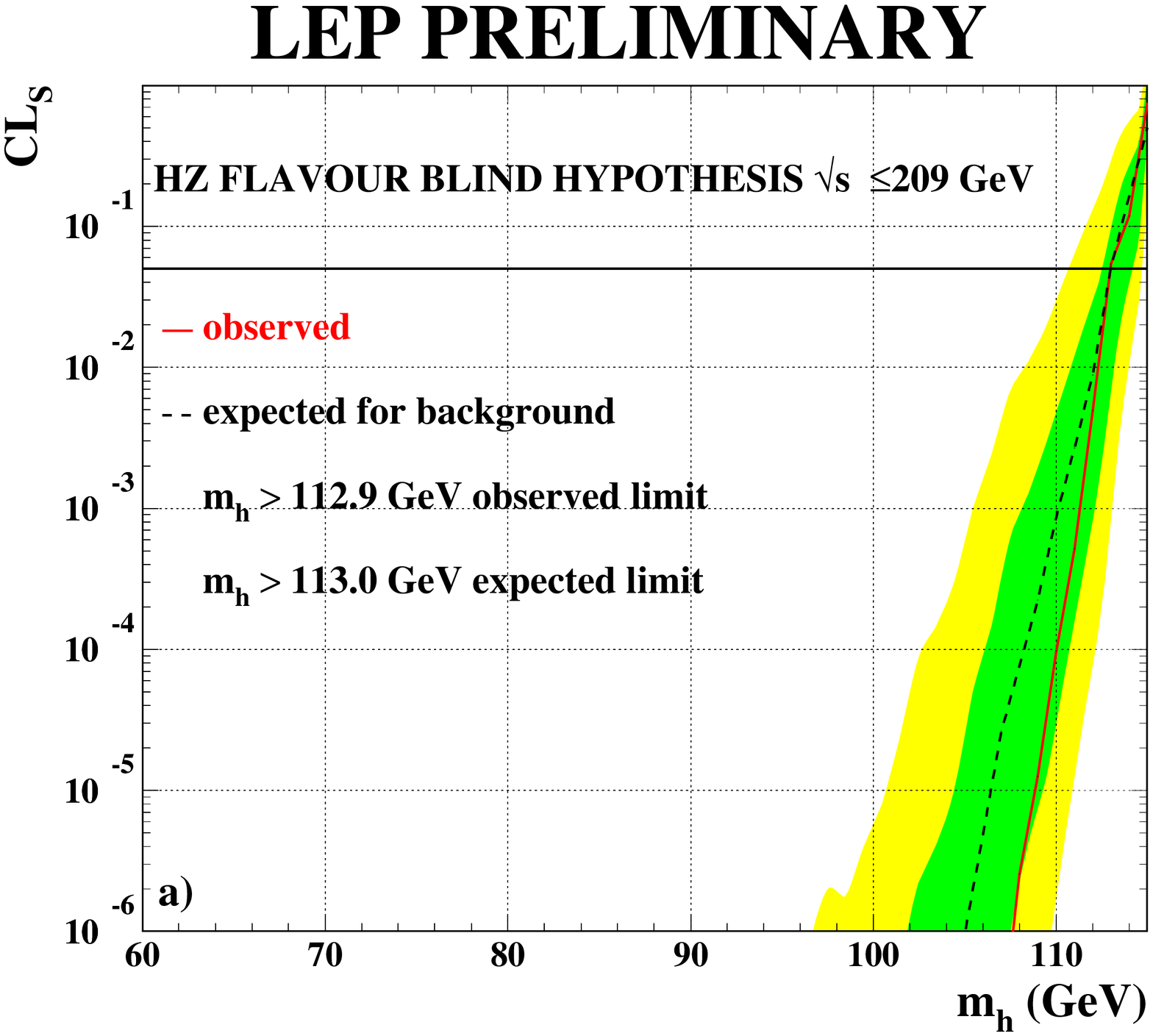}}    
   \mbox{\epsfxsize=13.0cm\epsfysize=9.0cm\epsffile{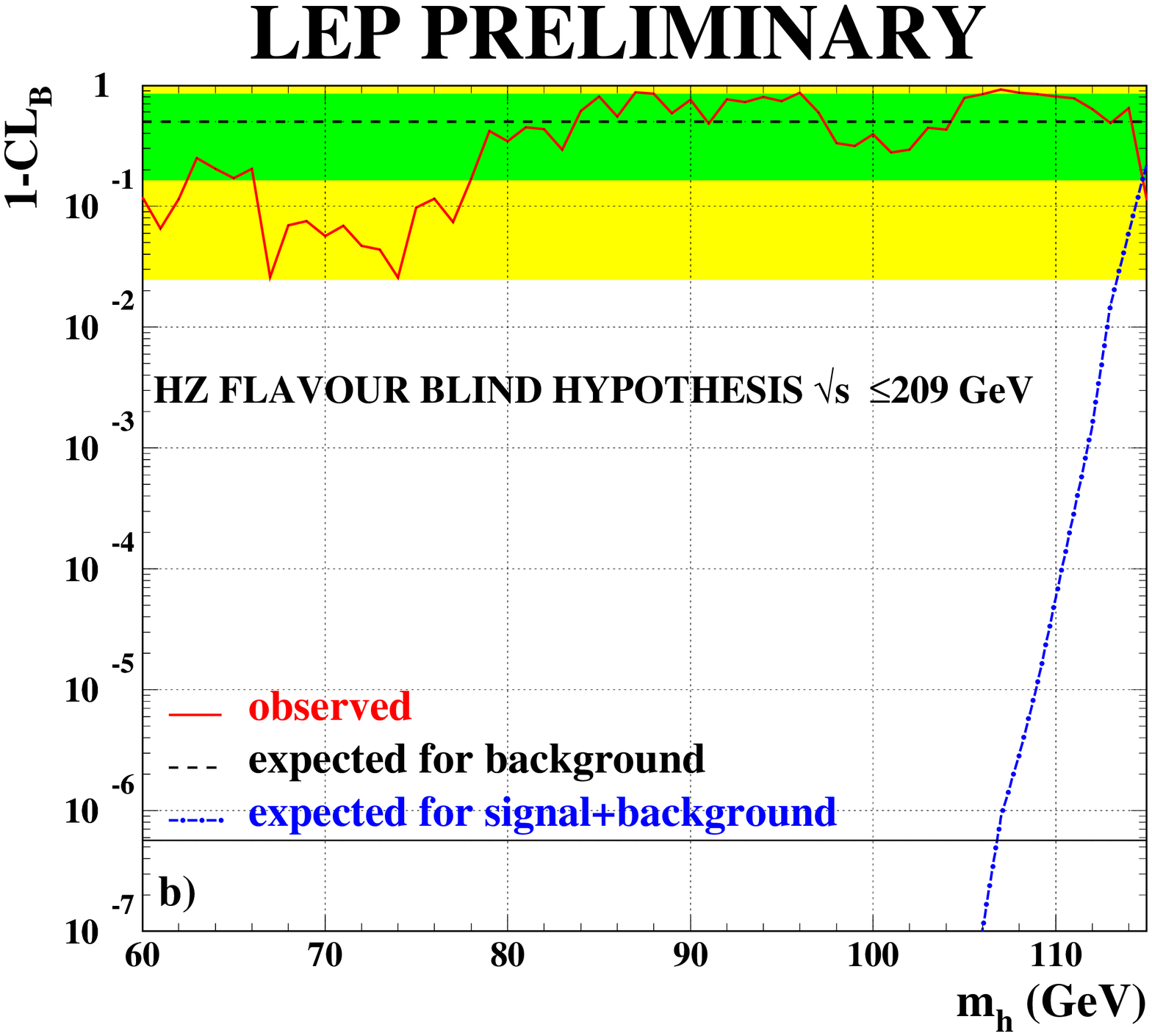}}  
  \caption{Combined LEP confidence levels as a function of the Higgs boson mass
in the flavour independent hypothesis, computed with the likelihood ratio 
technique~\cite{O-stat}, assuming production 
cross-sections equal to those in the Standard Model and
BR(h$\rightarrow$ hadrons) = 1.0.
The confidence levels for the signal and background-only hypotheses are 
shown in the upper (a) and lower (b) plots, respectively. The 
curves are the observed (solid) and expected median (dashed) confidences 
from background-only experiments, and the bands are the corresponding 
68.3 \% and 95 \% confidence intervals. In the lower plot, the dot-dashed line
shows the expected median confidence from experiments including an expected
signal of mass given in abscissa.}
  \label{fig:lepclsb}
  \end{center}
\end{figure}

The confidence levels 
$CL_s$ and $CL_b$ obtained from the full combination in the signal and 
background-only hypotheses are shown as a function of the mass in 
Figure \ref{fig:lepclsb}.a and \ref{fig:lepclsb}.b, respectively. 
Good overall agreement
can be seen between the observation and expectation in the absence of
a signal. The slightly depressed values for $1-CL_b$ for masses 
below 80 GeV/c$^2$, witnessing excesses of data in this region, 
may be the results of some statistical or 
systematic effect, and should be investigated further.
A 5 sigma discovery corresponds to a value of
5.7 $10^{-7}$, as indicated by the horizontal line. The maximal sensitivity
for such a discovery is reached for an assumed Higgs boson mass of 107
GeV/c$^2$, when this line is intersected by the expected median confidence in
the background-only hypothesis.

\begin{table}[htbp]
\begin{center}
\begin{tabular}{|c|c|c|} \hline
\hspace{0.5cm} Collaboration \hspace{0.5cm} & \hspace*{0.5cm} Obs. limit (GeV/c$^2$)\hspace*{0.5cm} & \hspace{0.5cm} Exp. median limit (GeV/c$^2$) \hspace*{0.5cm}\\ \hline
ALEPH      & 109.3     &  108.4   \\
DELPHI     & 109.6     &  108.8   \\
L3         & 111.6     &  109.3   \\
OPAL       & 109.4     &  108.5   \\ \hline \hline
{\bf LEP}  & 112.9     &  113.0   \\ \hline

\end{tabular}
\end{center}
\caption{Flavour independent observed and expected 95\% CL lower limits on 
the mass of the Higgs boson, assuming production 
cross-sections equal to those in the Standard Model and
BR(h$\rightarrow$ hadrons) = 1.0. The evaluation was performed 
using the likelihood ratio technique~\cite{O-stat}. Systematic uncertainties
were not included.}
 \label{leplim}
\end{table}

\begin{figure}[tbh]
  \begin{center}
   \mbox{\epsfxsize=15.0cm\epsfysize=15.0cm\epsffile{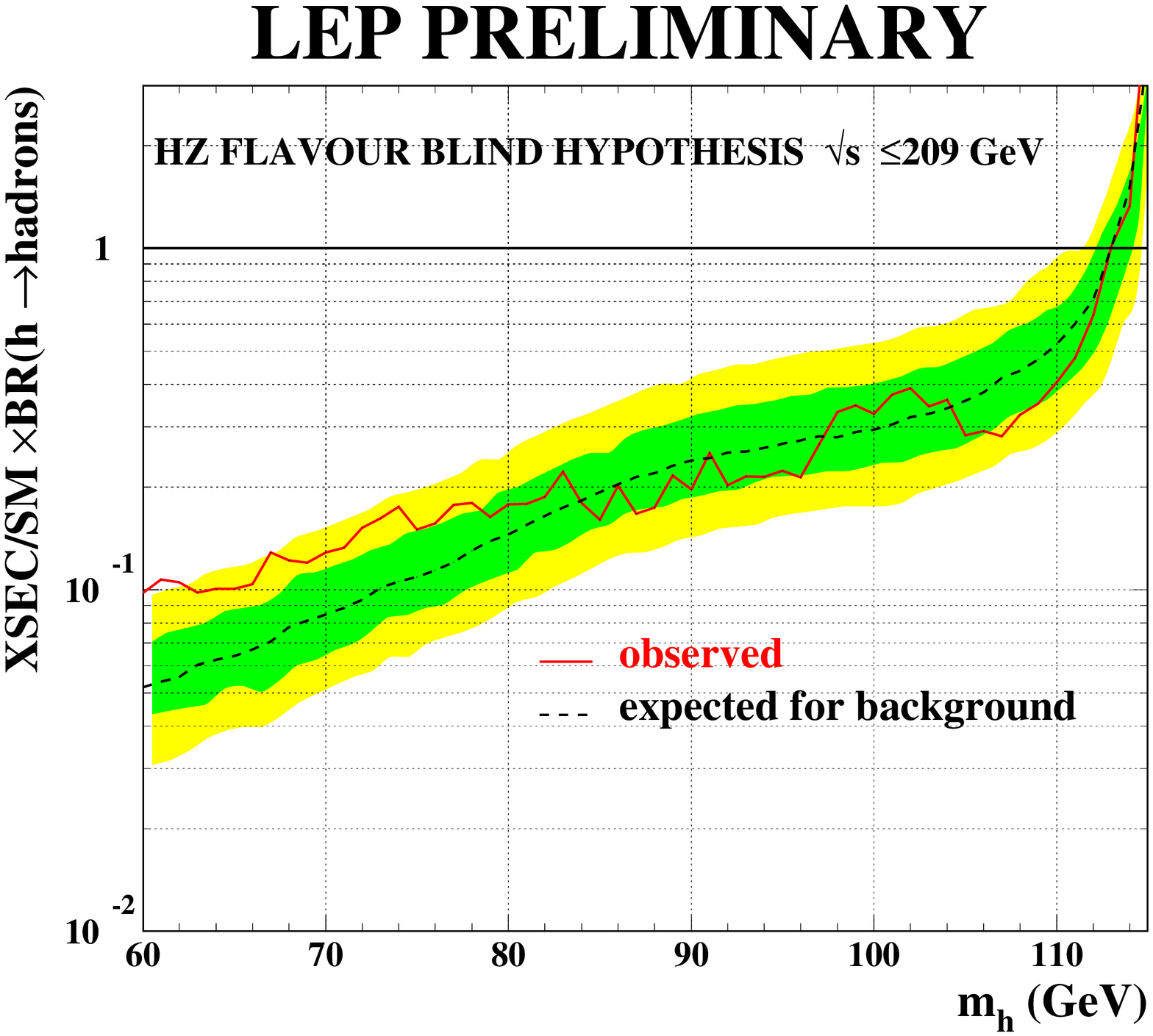}}    
  \caption{Combined flavour independent LEP observed and 
expected 95 \% CL upper limits on the production cross-section 
as a function of the Higgs boson mass, normalised
to the expected Standard Model values, computed assuming
BR(h$\rightarrow$ hadrons) = 1.0.
The computation was done with the likelihood ratio technique~\cite{O-stat}. 
The curves are the observed (solid) and expected median (dashed) 
excluded ratios, and the bands correspond to 68.3 \% and 95 \%
confidence intervals from the background-only experiments.}
  \label{fig:lepXSexcl}
  \end{center}
\end{figure}

Upper limits on the production cross-section as a function of mass
were also determined for each individual collaboration, and 
combined using the data from all four collaborations. Typically, 
cross-sections larger than about 10-60 \% of the expected
Standard Model value were excluded at 95\% CL in the mass range 60-100 GeV/c$^2$,
by each collaboration alone. The combined exclusion from the 
four collaborations is shown as a function of the mass in 
Figure \ref{fig:lepXSexcl}. In the same mass range, it is
possible to exclude cross-sections larger than about 
a few to 30 \% of the expected Standard Model value with the full
LEP-2 data set. Good overall agreement between the observation and 
expectation can be seen also here, except in the lower mass range,
where the weaker observed limit is resulting from some excesses of data 
in this region, as was already noted.


\section*{Acknowledgements}

We are greatly indebted to our technical 
collaborators, to the members of the CERN-SL Division for the excellent 
performance of the LEP collider, and to the funding agencies for their
support in building and operating the ALEPH, DELPHI, L3 and OPAL detectors. 
We also acknowledge the fruitful collaboration between all four collaborations,
and the open spirit which prevailed and enabled to implement efficiently the
established combination procedures to the new channels described in this note.

\pagebreak

\clearpage

\clearpage

\end{document}